# DDoS attack detection method based on feature extraction of deep belief network

Li Yijie, Zhai Shang, Chen Mingrui

**Absrtact:** Distributed Denial of Service (DDOS) attack is one of the most common network attacks. DDoS attacks are becoming more and more diverse, which makes it difficult for some DDoS attack detection methods based on single network flow characteristics to detect various types of DDoS attacks, while the detection methods of multi-feature DDoS attacks have a certain lag due to the complexity of the algorithm. Therefore, it is necessary and urgent to monitor the trend of traffic change and identify DDoS attacks timely and accurately. In this paper, a method of DDoS attack detection based on deep belief network feature extraction and LSTM model is proposed. This method uses deep belief network to extract the features of IP packets, and identifies DDoS attacks based on LSTM model. This scheme is suitable for DDoS attack detection technology. The model can accurately predict the trend of normal network traffic, identify the anomalies caused by DDoS attacks, and apply to solve more detection methods about DDoS attacks in the future.

# 1. introduction

Distributed Denial of Service (DDoS) attacks refer to the combination of multiple computers as attack platforms to launch DDoS attacks on one or more targets with the help of client/server technology, thus doubling the power of denial of service attacks. Among them, DDoS can be divided into the following attack modes: (1) disturbing or even blocking normal network communication by overloading the network; (2) overloading the server by submitting a large number of requests to the server; (3) blocking a user's access to the server; (4) blocking the communication between a service and a specific system or individual. With the increasing popularity and development of the Internet, the relationship between the Internet and people's daily life is getting closer and closer. But the Internet is a double-edged sword. It brings us convenience, but it also brings us many problems. Among many problems, network security is the most important one. At present, the frequency of network intrusion is getting higher and higher, and the harmfulness of intrusion is getting greater and greater, especially the intrusion that consumes network resources is becoming more and more intense. As a valuable resource, network bandwidth directly affects the quality of people's access to the network. Therefore, how to ensure the effective use of bandwidth resources, timely detection and defense of malicious consumption of network bandwidth behavior is an important research direction.

At present, the research on DDoS mainly focuses on three aspects: prevention, detection and tracking. The first line of defense against DDoS attacks is attack prevention. The purpose of prevention is to take measures to prevent the attacker from launching DDoS attacks and endangering the network when the attack has not yet occurred. When an attack does occur, it needs to be responded. The purpose of response tracking is to eliminate or mitigate attacks and minimize the harm to the network caused by attacks. Response tracking research can be divided into tracking when an attack occurs and tracking after an attack occurs. In order to respond to the attack as soon as possible, it is necessary to detect the existence of the attack as soon as possible.

The purpose of attack detection is to identify attack data in network flow. For general DDoS attack detection, firstly, feature is extracted from network data stream, secondly, appropriate model is selected to predict unknown data stream, and finally the prediction results are evaluated and further optimized.

Compared with other attacks, DDoS attacks are very similar to the normal background traffic, which makes the eigenvalues of samples have great uncertainty and fuzziness, and has a great impact on the classification effect. Therefore, it is very important to select a suitable classification feature.

In order to solve the above problems, a method of DDoS attack detection based on deep belief network feature extraction and LSTM model is proposed in this paper. Firstly, this method extracts IP packet features using deep belief network, then establishes LSTM traffic prediction model, and finally identifies DDoS attacks based on the established LSTM model. This scheme is suitable for DDoS attack detection technology. The model can accurately predict the trend of normal network traffic, identify the anomalies caused by DDoS attacks, and apply to solve more detection methods about DDoS attacks in the future.

## 2. DDoS Attack Characteristics

### 2.1 DDoS Attack Characteristics

With the development of attack technology, single DDoS attack data packet is usually legitimate in protocol and content. Various DDoS attack software can be downloaded from the Internet at will, so any Internet user may become a potential threat to network security. DDoS attacks are becoming more and more easy to implement, wide-ranging, destructive, complex, dynamic, difficult to resist and track.

### 2.2 DDoS Attack Feature Extraction

Based on the analysis of DDoS attack characteristics, this paper proposes a DDoS feature extraction method based on deep belief network to extract features.

Deep belief network is a method of learning model with deep structure proposed by Hinton [2,3]. It is also one of the first non-convolutional models to successfully apply deep architecture training. Compared with the traditional neural network, it has stronger modeling and representation ability, so it can complete complex function approximation. Depth belief network can be seen as a series of constrained Boltzmann machine modules stacked together [4]. The front layer is the visible layer of the next hidden layer and the input of the next hidden layer. Restricted Boltzmann machine is an undirected probability graph model with one layer of observable variables and one layer of latent variables. It is also an energy-based model and a recursive neural network.

The confined Boltzmann machine structure is shown in Figure 1.

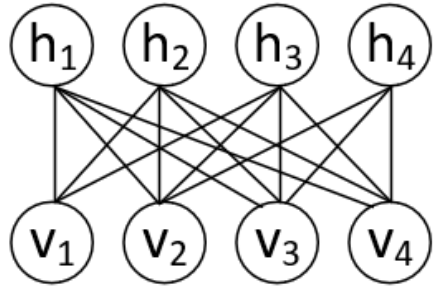

Figure 1. RBM composition diagram

It is shown as a bipartite graph. RBM has only two layers of neurons. One layer is called the display layer, which is used to input training data. One layer is called the hidden layer, which is used as a feature detector. There is a connection between layers, but there is no connection between units in the layer. The explicit layer is used to input training data, and the hidden layer is trained to capture the correlation of higher-order data in the visual layer. The neurons in the restricted Boltzmann machine are Boolean, i.e. they can only take two states: 0 and 1, state 1 means activation, and state 0 means inhibition.

Depth belief network can be seen as a series of confined Boltzmann machine modules stacked. [4] Its structure is shown in Figure 2.

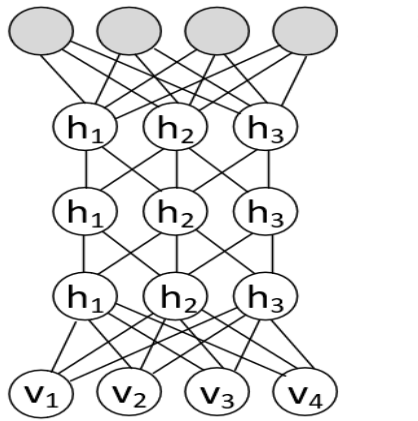

Figure 2. DBN composition diagram

The energy function E is defined as:

$$E(v, h; \theta) = -\sum_{i=1}^{V}\sum_{j=1}^{H} w_{ij} v_i h_j - \sum_{i=1}^{V} b_i v_i - \sum_{j=1}^{H} a_j h_j$$

$$E(v, h; \theta) = -\sum_{i=1}^{V}\sum_{j=1}^{H} w_{ij} v_i h_j - \frac{1}{2}\sum_{i=1}^{V} (v_i - b_i)^2 - \sum_{j=1}^{H} a_j h_j$$

$\theta = \{w, a, b\}$, $w_{ij}$ is the connection weight of the visual unit and the implicit element. $b_i$ and $a_j$ is corresponding to the offset. $v_i$ is the number of visible units, $h_j$ is the number of hidden units.

## 3. Detection of DoS attacks based on LSTM

Long-term and short-term memory model [7] is a special RNN [8,9] model, which is proposed to solve the problem of gradient dispersion of RNN model. Among them, the internal state of RNN network can show the dynamic sequential behavior. Unlike feedforward neural networks, RNN can use its internal memory to process input sequences of arbitrary time series, which makes it easier to process such as non-segmented handwriting recognition, speech recognition and so on. However, RNN has two problems, namely, gradient disappearance and gradient explosion. The original intention of LSTM design is to solve the problem of long-term dependence in RNN, so that remembering long-term information becomes the default behavior of neural network, rather than a lot of effort to learn. LSTM model replaces RNN cells in the hidden layer with LSTM cells to make them have long-term memory ability. At present, the popular LSTM model structure is Cho, et al [8], which uses the forgetting gate and the input gate to receive and input parameters respectively. In the LSTM neural network model, the forward calculation method can be expressed as:

$$f_t = \sigma(W_f[h_{t-1}, X_t] + b_f)$$
$$i_t = \sigma(W_i[h_{t-1}, X_t] + b_i)$$
$$\widetilde{C}_t = tanh(W_c[h_{t-1}, X_t] + b_c)$$
$$C_t = f_t * C_{t-1} + i_t * \widetilde{C}_t$$
$$o_t = \sigma(W_o[h_{t-1}, X_t] + b_o)$$
$$h_t = o_t * \tanh(C_t)$$

Among them, f, i, c, o respectively represent forgetting gate, output gate, cell state, output gate, W and B are corresponding weight coefficient matrix and bias term. Rou and tanh are sigmoid and hyperbolic tangent activation functions respectively. From the introduction of the formula, it can be seen that LSTM can solve the long-term dependence problem in RNN. Therefore, a neural network prediction model based on LSTM can be established, and the extracted features can be used for DDoS attack recognition and detection.

## 4 Conclusion

For the research of DDoS attack detection, this paper proposes a method of DDoS attack detection based on deep belief network feature extraction and LSTM model. Firstly, this method extracts IP packet features using deep belief network, then establishes LSTM traffic prediction model, and finally identifies DDoS attacks based on the established LSTM model. This scheme is suitable for DDoS attack detection technology. The model can accurately predict the trend of normal network traffic, identify the anomalies caused by DDoS attacks, and apply to solve more detection methods about DDoS attacks in many fields [11] in the future.